# Microscopic analysis of the chemical reaction between Fe(Te,Se) thin films and underlying $CaF_2$

A. Ichinose,[1,*] I. Tsukada,[1] M. Hanawa,[1] Seiki Komiya,[1]

T. Akiike,[2] F. Nabeshima,[2] Y. Imai,[2] A. Maeda [2]

[1] *Central Research Institute of Electric Power Industry,*

*2-6-1 Nagasaka, Yokosuka, Kanagawa 240-0196, Japan*

[2] *Department of Basic Science, The University of Tokyo,*

*3-8-1 Komaba, Meguro-ku, Tokyo 153-8902, Japan*

**Abstract**

To understand the chemical reaction at the interface of materials, we performed a transmission electron microscopy (TEM) observation in four types of Fe(Te,Se) superconducting thin films prepared on different types of substrates: $CaF_2$ substrate, $CaF_2$ substrate with a $CaF_2$ buffer layer, $CaF_2$ substrate with a FeSe buffer layer, and a $LaAlO_3$ substrate with a $CaF_2$ buffer layer. Based on the energy-dispersive X-ray spectrometer (EDX) analysis, we found possible interdiffusion between fluorine and selenium that has a strong influence on the superconductivity in Fe(Te,Se) films. The chemical interdiffusion also plays a significant role in the variation of the lattice parameters. The lattice parameters of the Fe(Te,Se) thin films are primarily determined by the chemical substitution of anions, and the lattice mismatch only plays a secondary role.

**I. Introduction**

Following the discovery of superconductivity in iron pnictide materials,[1] a considerable number of experiments have been performed to fabricate thin films and single crystals to pursue higher critical temperatures, $T_C$. These new superconductors have been categorized into five major families: the '1111 family' (the first material, LaFeAsO$_{0.89}$F$_{0.11}$[1] and its derived compounds); the '111 family', including LiFeAs,[2] the '122 family', including BaFe$_2$As$_2$,[3] the '11 family', including FeTe$_{1-x}$Se$_x$,[4] and the 'perovskite family', including Sr$_2$ScO$_3$FeAs with perovskite structures.[5] Among these families of materials, the '11 family' has the simplest structure, which is an advantage for practical applications. Although the $T_C$ of bulk FeSe is rather low, approximately 8 K, it exhibits a strong pressure dependence,[6] and was reported to exceed 37 K under high pressure, approximately 7 GPa.[7,8] Therefore, applying a strain using a lattice mismatch between the substrates and thin films of these superconducting materials is of particular interest to researchers. E. Bellingeri *et al.* reported that the dependence of the $T_C$ on the thickness of the film is explained by the change of the strain effect, and they obtained the highest $T_C$ of 21 K, which provides evidence that a strain can change the $T_C$.[9] Mele *et al.*, have reported that the upper critical field, $B_{C2}\|(0)$, can be as high as 60 T, even for a low-$T_C$ Fe(Te,S) film with a $T_C$ of 4 K,[10] which indicates that the "11"-superconducting films have sufficient potential for low-temperature applications.

In early studies on thin-film growth, conventional oxide substrates were widely used for growing '11' films. Several research groups have reported that the $T_C$ of the film strongly depends on the oxide substrate materials.[11-15] We observed that the $T_C$ of the '11' film strongly depends on the amount of oxygen that penetrates into the film from the oxide substrates.[16,17] Based on these findings, we have proposed the use of CaF$_2$ as a substrate and have succeeded in improving the $T_C$ and other parameters.[18] One film exhibits an onset of superconductivity at 16.6K and zero resistivity at 15.7 K. Note that in the case of using CaF$_2$ substrates, the *a*-axis of the FeTe$_{0.5}$Se$_{0.5}$ superconducting film becomes shorter and the *c*-axis of the film becomes longer than those in the bulk crystals.[19] Surprisingly, this lattice deformation cannot be explained by a conventional epitaxial-strain effect because the length of the *a*-axis in CaF$_2$ is considerably longer than that in bulk FeTe$_{0.5}$Se$_{0.5}$. To determin an appropriate mechanism for the lattice deformation when using CaF$_2$ substrates for the "11"-superconducting films, we prepared Fe(Te,Se) thin films on various substrates and analyzed the microstructures at the interface using transmission electron microscopy (TEM).

**II. Experimental Procedures**

The thin films were deposited using the pulsed-laser deposition (PLD) method with a KrF excimer laser ($\lambda$=248 nm) using polycrystalline targets with a nominal composition of Fe : Se : Te = 1.0 : 0.4 : 0.6. Buffer layers were also deposited using polycrystalline CaF$_2$ and FeSe target. The deposition temperatures for Fe(Te,Se) and FeSe were 280ºC and for CaF$_2$ was 340ºC. The repetition frequency and energy of the laser were 10 Hz and 300 mJ, respectively. The film preparation has been described in detail elsewhere.[15,17] The in-plane and out-of-plane crystal orientations were characterized by X-ray diffraction as Fe(Te,Se) (001) // CaF$_2$ (100) or LaAlO$_3$ (100). The temperature dependence of the resistivity was measured using the standard four-probe method.

We prepared four Fe(Te,Se) thin films on different substrates. Table I summarizes the specifications of the four films. Film A is a FeTe$_{0.5}$Se$_{0.5}$ film grown on a bare CaF$_2$ (100) substrate, which is identical to sample A in Ref. 18. Films B and C are FeTe$_{0.6}$Se$_{0.4}$ films grown on a CaF$_2$ (100) substrate with CaF$_2$ and FeSe inserted between the film and the substrate, respectively. Film D is a FeTe$_{0.6}$Se$_{0.4}$ film grown on a LaAlO$_3$ (100) substrate with an inserted CaF$_2$ buffer layer.

The samples for TEM analyses were prepared by cutting and milling using a focused ion beam (FIB), the so-called micro-bridge sampling technique. The microstructures of these samples were examined using a JEOL TEM-2100F microscope with an energy-dispersive X-ray spectrometer (EDX). Nanobeam electron diffractions (NBD) experiments were performed to evaluate the local structures in the interfaceial area between the films and the substrates. An incident electron beam was theoretically focused to a spot with a diameter of 1 nm in the NBD experiments.

**III. Results and Discussion**

Figure 1 shows the temperature dependence of the resistivities for films A, B, C, and D. These films exhibit an onset of superconductivity, $T_C^{onset}$, at 16.6, 13.7, 14.4 and 13.0 K, respectively. The $T_C^{onset}$ of film A exhibits the highest $T_C$, which is reasonable because this film has a different Se/Te ratio. The $T_C^{onset}$ of the other films are approximately 14 K, which is not a considerably low value for $FeTe_{0.6}Se_{0.4}$.

In the following several paragraphs, we will present the details of the TEM observation of each film in detail. A cross-sectional scanning TEM (STEM) image of film A is presented in Fig. 2 (a). A high-resolution cross-sectional TEM image of film A, which corresponds to the area indicated in Fig. 2 (a), and a selected-area electron diffraction (SAED) pattern are presented in Figs. 2 (b) and (c), respectively. A yellow dashed line is drawn at the interface in Fig. 2 (a). A bright layer that was approximately 5 nm in width extended from the interface between the $CaF_2$ substrate and the superconducting film. This region is not amorphous, as confirmed by SAED pattern. According to Fig. 2 (c), $FeTe_{0.5}Se_{0.5}$ [110] // $CaF_2$ [100] and $FeTe_{0.5}Se_{0.5}$ (001) // $CaF_2$ (001) and that lengths of the *a*- and *c*-axes of film A are 0.376 and 0.594 nm, respectively.

A cross-sectional STEM image of film B is shown in Fig. 3 (a). The interface between $FeTe_{0.6}Se_{0.4}$ and the $CaF_2$ buffer layer exhibits a peculiar structure. There are many triangular shapes that are several tens of nm in length at this interface, which indicates that the structure of this type of surface overlaps along the incident direction of the electron beams. Surprisingly, the surface of the superconducting film recovers excellent flatness, even though the surfaces of the underlying layers are extremely rough. Therefore, the "11" superconductors can grow two-dimensionally irrespective of the degree of surface flatness of an underlying layer. A cross-sectional, high-resolution TEM image near the two interfaces of film B, which corresponds to the square areas I and II shown in Fig. 3 (a), are shown in Figs. 3 (b) and (c), respectively. Figure 3 (d) presents the SAED pattern at the interface between the $CaF_2$ buffer layer and the $FeTe_{0.6}Se_{0.4}$ superconducting film. The $CaF_2$ buffer layer and the $CaF_2$ substrate are well ordered due to a homoepitaxial growth. The interface between the $FeTe_{0.6}Se_{0.4}$ superconducting film and the $CaF_2$ buffer layer is more complicated. The yellow dashed lines are drawn perpendicular to the equivalent direction of $CaF_2$ [111] in Fig. 3 (c), which indicates that the plane corresponding to the yellow dashed lines is equivalent in $CaF_2$ (111). In other words, the pyramidal structures surrounded by the (111) planes are observed as an overlapped image. The preferential crystal growth direction of $CaF_2$ is generally the [110] direction.[20] In this case, the pyramidal structure surrounded by the (111) planes is known to be formed when $CaF_2$ has the c-axis orientation. Note that the 'bright" layer observed at the interface of film A is not detected in film B. Although a thin reaction layer exists at the interface, the surface irregularities greater than approximately 10 nm in height conceals a thin layer that is approximately 5 nm in thickness. According to the SAED pattern, $FeTe_{0.6}Se_{0.4}$ [100] // $CaF_2$ [110] and $FeTe_{0.6}Se_{0.4}$ (001) // $CaF_2$ (001), and the lengths of the *a*- and *c*-axes of film A are 0.377 and 0.605 nm, respectively. Another interesting finding is that the $FeTe_{0.6}Se_{0.4}$ layer begins to grow immediately above the bottom of the valley of the $CaF_2$ buffer layers.

This result suggests that the growth direction of Fe(Te,Se) is also strongly governed by the (111) facet of $CaF_2$. This result suggests the presence of an intimate epitaxial relationship between $CaF_2$ and Fe(Te,Se) and suggests the possibility of growing non-c-axis oriented 11 films.

A cross-sectional STEM image of film C is shown in Fig. 4 (a). Although the interface between the FeSe buffer layer and the $FeTe_{0.6}Se_{0.4}$ superconducting film is not clear, slight but detectable differences in the contrast change in each layer can distinguish the two layers. A cross-sectional, high-resolution TEM images near the two interfaces of film C, which correspond to the square areas I and II shown in Fig. 4 (a), are shown in Figs. 4 (b) and (c), respectively. Figure 4 (d) shows the SAED pattern at the interface between the FeSe buffer layer and the $FeTe_{0.6}Se_{0.4}$ superconducting film. Each interface is indicated with yellow dashed lines in Figs. 4 (b) and (c). A bright, thin layer with a thickness of approximately 5 nm is observed in the side of the $CaF_2$ substrate, which is the same as with film A. The same reaction is believed to occur at the interface of the $CaF_2$ substrate in both films of A and C. The interface between the FeSe and the $FeTe_{0.6}Se_{0.4}$ is well ordered, which is reasonable because they have almost the same crystal structure. The high-resolution TEM images of the FeSe buffer layer and the $FeTe_{0.6}Se_{0.4}$ superconducting layer, which correspond to the square areas of III and IV in Fig. 4(a), are shown in Figs. 5 (a) and (b), respectively. The crystal structure of $FeTe_{0.6}Se_{0.4}$ is more ordered than that of FeSe. The crystal structures of $FeTe_{1-x}Se_x$ at x > 0.8 and x < 0.8 possess a similar tetragonal lattice, but a discontinuous change in the lattice lengths is observed near x~0.8.[21] This difference in the crystal structure appears to affect the morphology of the materials in the nm range. According to the SAED pattern, $FeTe_{0.6}Se_{0.4}$ [100] // FeSe [100] and $FeTe_{0.6}Se_{0.4}$ (001) // FeSe (001), and the lengths of the *a*- and *c*-axes of the FeSe buffer layer and of the $FeTe_{0.6}Se_{0.4}$ superconducting layer are 0.371 and 0.551 nm and 0.375 and 0.604, respectively.

Furthermore, the spots in the pattern from FeSe are slightly more dispersed than those from FeTe$_{0.6}$Se$_{0.4}$, which is consistent with the TEM results, as shown in Figs. 5 (a) and (b).

A cross-sectional STEM image of film D is shown in Fig. 6 (a). A thick, bright layer homogeneously exists on the top of the LaAlO$_3$ substrate. The peculiar pyramidal structure of the CaF$_2$ buffer layer is almost the same as that of Film B. The surface of the superconducting film on the rough CaF$_2$ buffer layer is flat. A cross-sectional high-resolution TEM images near the two interfaces of film D, which is correspond as a square area in I and II in Fig. 6(a), are shown in Figs. 6 (b) and (c), respectively. Figures 6 (d) and (e) show the SAED patterns at the interface between the LaAlO$_3$ substrate and the CaF$_2$ buffer layer and between the CaF$_2$ buffer layer and the FeTe$_{0.6}$Se$_{0.4}$ superconducting film. Each interface is represented as yellow dashed lines in Figs. 6 (b) and (c). In Fig. 6 (b), a wide bright region with a thickness greater than 10 nm is observed inside of the LaAlO$_3$ substrate. To evaluate the local crystallographic structure at this bright area, we performed a nanobeam electron diffraction measurement in the region represneted as a circle in III. Figure 7 shows the diffraction pattern from area III, in which clear diffraction spots are no longer observed and a halo ring appears. This behavior indicates that the bright area becomes an amorphous-like structure. Because the CaF$_2$ buffer layer is epitaxially grown on the LaAlO$_3$ substrate, it is concluded that the amorphous region at the interface begins forming after a sufficient nucleation of CaF$_2$ has concluded. The yellow dashed lines in Fig. 6 (c) correspond to the CaF$_2$ (001) and the CaF$_2$ (111). The surface structure of the CaF$_2$ buffer layer has a pyramid surrounded by CaF$_2$ (111) facets, which is the same as film B. Therefore, the bright thin layer at the surface of the CaF$_2$ buffer layer is invisible for the same reason as film B, even though a similar layer might be present at the surface. According to the SAED patterns, FeTe$_{0.6}$Se$_{0.4}$ [100] // CaF$_2$ [110] // LaAlO$_3$ [100] and FeTe$_{0.6}$Se$_{0.4}$ [001] // CaF$_2$ [001] // LaAlO$_3$ [100], and the lengths of the *a*- and *c*-axes of Film D are 0.375 and 0.605 nm, respectively.

Figures 8 (a)-(d) present the EDX results for films A-D, respectively. Figure 8 (e) presents the EDX result of the expanded area near the substrate interface for film C. In all figures, the amount of fluorine in CaF$_2$ decreases closer to the interface. In Fig. 8 (b), although the amount of fluorine is almost constant in the substrate, the amount of fluorine is decreasing as the interface of the superconducting film is approached. In Fig. 8 (d), which presents a similar situation, the amount of fluorine decreases as both interfaces of the CaF$_2$ buffer layers are approached. Specifically, at the interface between the CaF$_2$ buffer layer and the LaAlO$_3$ substrate, the presence of fluorine is clearly detected, but calcium is not detected in the amorphous layer at the top region of the substrate. Although calcium fluoride is known to be a stable material, the present result indicates that the fluorine ions can easily move at the relatively low temperature of 340ºC. Furthermore, the fluorine ions appeared to move more easily in the post-deposited CaF$_2$ buffer layer than in the CaF$_2$ substrates. This difference might be due to different surface conditions (the CaF$_2$ (111) facet is dominant in the CaF$_2$ buffer layer, whereas the CaF$_2$ (100) surface is dominant in the CaF$_2$ substrate), and/or the difference in the density of defects between them.

In the case of the oxide substrates, the reaction layers definitely existed in the superconducting films, and the layer was considered to be an amorphous-like layer based on a TEM image of a thicker reaction layer.[16,17] In other words, the oxygen penetrates into the film and perturbs the crystal structure of the superconducting film. The penetration of oxygen into Fe(Te,Se) induces significant disorder in the crystal structure of Fe(Te,Se), and it is thus easily detected by TEM observation. In contrary, in the case of the fluoride substrates, the reaction layers are always observed inside of the CaF$_2$ and never found in the superconducting films. It is difficult to precisely determine the amount of fluorine ions in the superconducting films due to the lower detection accuracy for fluorine ions than any other elements. However, according to the experimental results shown in Fig. 8, it is suggested that the fluorine ions not only penetrate from the CaF$_2$ substrates into the superconducting layers but are also incorporated into Fe(Te,Se). Two possible scenarios for the incorporation of fluorine ions are compared. One scenario is the substitution of the selenium ions with the fluorine ions and the other scenario is the intercalation of fluorine ions between the layers. In the former scenario, the selenium ions would appear to slightly penetrate the superconducting films into the CaF$_2$ substrate. In this case, the substitution of fluorine ions for the selenium ions likely occurs. At the present stage, we have no clear answer for the more probable scenario, but the study of F-substitution effects would be an interesting subject for future investigation.

The $T_C^{onset}$ of the Fe11-based superconductors is known to be correlated to the ratio of the lattice lengths, $c/a$.[16] The lengths of the $a$- and $c$-axes of the superconducting films on the four types of substrates were evaluated from the SAED patterns, and the results are summarized in Table I. Figure 9 shows the dependence of the $T_C^{onset}$ values on the lengths of the $a$- and $c$-axes as functions of $c/a$. The previous data for the FeTe$_{0.5}$Se$_{0.5}$ thin films on oxide substrates are also plotted in Fig. 9. The overall tendencies are similar to those discussed from the results of x-ray diffraction studies.[17] When the CaF$_2$ buffered substrates are used, the values of $c/a$ increase, the $a$-axis length becomes short and the $c$-axis length becomes long.  This result cannot be explained by a simple mismatch strain induced from the differences of lattice lengths between the superconducting film and the CaF$_2$ substrate. The length of the $a$-axis of a bulk superconductor is 0.3798 nm and the length of the CaF$_2$ crystal corresponding to the $a$-axis length of superconductors is 0.386 nm, which would instead induce a tensile strain into the superconducting films. However, the experimental results reveal that the length of the *a*-axis is in the range of 0.372-0.382 nm. In other words, the superconducting film shrinks without experiencing a tensile stress. Furthermore, the lengths of the *a*- and *c*-axes in the FeTe$_{1-x}$Se$_x$ monotonically increase with increasing Te content at the range of x<0.8.[21] When comparing FeTe$_{0.5}$Se$_{0.5}$ films on the CaF$_2$ substrate (including film A) and FeTe$_{0.6}$Se$_{0.4}$ films (films B-D), it is obvious that the length of the *a*-axis of the FeTe$_{0.6}$Se$_{0.4}$ films are almost constant and somewhat shorter than that of the FeTe$_{0.5}$Se$_{0.5}$ films. This result cannot be also explained by increasing Te content.

In the case of high-$T_C$ cuprate superconductors with a layered structure, the length of the *c*-axis depends on the electrostatic force between the layers. Consequently, the length of the *c*-axis increases when the electrostatic force is decreased through the substitution of a smaller valence ion or through the presence of vacancies.[22] In contrast, the length of the *a*-axis is not affected by the electrostatic force because of the rigid structure of the in-plane region of the structure, but it is affected by the ionic radius. The Fe-based superconductor has a layered structure similar to that of the high-$T_C$ cuprate superconductors. Based on the assumption that fluorine ions substitute for selenium ions, the length of the *c*-axis should increase when ions with a smaller absolute valence are substituted, and the length of the *a*-axis should decrease when ions with a smaller ionic radius are substituted. The radii of Se$^{2-}$ and F$^-$ are 0.184 and 0.119 nm, respectively, which indicates that the shorter length of the *a*-axis and the longer length of the *c*-axis of the superconducting films on CaF$_2$ substrates, when compared to those on oxide substrates, can be explained by the partial substitution of F$^-$ ions for Se$^{2-}$ ions.

In the case of the slight increase of Te content, the FeTe$_{0.6}$Se$_{0.4}$ superconducting films using CaF$_2$ buffer layers have almost the same width and rather shorter *a*-axis lengths compared to that of the FeTe$_{0.5}$Se$_{0.5}$ superconducting films on the CaF$_2$ substrates. This result may be explained by considering the fact that the CaF$_2$ buffer layer has a columnar structure and a rough structure. This morphology makes the ion diffusion easier because of a considerable number of passages along the grain boundaries and a wide contact area. Consequently, the change of the *a*-axis length is determined by a delicate balance of two opposite effects: one is the decrease of the length of the *a*-axis resulting from $F^-$ (0.119 nm) substitution for $Se^{2-}$ (0.184 nm) and the other is the increase of the length of the *a*-axis by $Te^{2-}$ (0.207 nm) substitution for $Se^{2-}$. At this time, the effect of $F^-$ substitution appears to be predominant because of the large difference between the ionic radius of $F^-$ and $Se^{2-}$.

The critical temperature has a linear dependence on the length of the *a*-axis[9] or the ratio of *c*/*a*.[16] The data in the present experiment may not need to follow this rule due to the different contents of Te. However, according to the above discussion, the possibility of the substitution of $F^-$ for $Se^{2-}$ is suggested. Here, it is emphasized that the substitution of $F^-$ for $Se^{2-}$ refers to electron doping into the superconductors, which may be demonstrated using several experimental techniques. In this paper, the possibility of the fluorine substitution is discussed only from the view point of crystal structures. The superconductivity of the films on the $CaF_2$ substrates should be discussed while considering the effects of both the change in the lattice length and electron doping.

**IV. Conclusions**

The four types of substrates used to grow Fe(Te,Se) superconducting films by PLD, their microstructures and the ion distributions across the interface were analyzed using TEM. The Fe11-based superconductors grow two-dimensionally regardless of the morphology of the underlying layers. Furthermore, the fluorine ions are observed to easily move during film depositions. The possibility of fluorine ions from the $CaF_2$ substrates penetrating into the Fe(Te,Se) films is indicated not only by EDX analysis but also by a consideration of the lattice parameters estimated from the SAED experiments.

The change of the lattice lengths of the superconducting films on $CaF_2$ substrates is also consistent with the scenario of the partial substitution of selenium ions with fluorine ions. The substitution of selenium ions with fluorine ions likely works as electron doping. However, further investigations are required to reveal why the $CaF_2$ substrates are more suitable than oxide substrates for the growth of iron chalcogenide films. The electronic structure of the superconducting films on $CaF_2$ substrates will be investigated in the near future.

**Acknowledgements**

This research was supported by the Strategic International Collaborative Research Program (SICORP), Japan Science and Technology Agency.

**References**

* Electronic address: ai@criepi.denken.or.jp

[1] Y. Kamihara, T. Watanabe, M. Hirano, and H. Hosono, J. Am. Chem. Soc. **130**, 3296 (2008).

[2] X.C. Wang, Q.Q. Liu, Y.X. Lv, W.B. Gao, L.X. Yang, R.C. Yu, F.Y. Li, and C.Q. Jin, Solid State Commum. **148**, 538 (2008).

[3] M. Rotter, M. Tegel, D. Johrendt, I. Schellenberg, W. Hermes, and R. Pottgen, Phys. Rev. B **78**, 020503 (2008).

[4] F.S. Hsu *et al.*, Proc. Natl. Acad. Sci. USA **105**, 14262 (2008).

[5] H. Ogino, Y. Matsumura, Y. Katsura, K. Ushiyama, S. Horii, K. Kishio, and J, Shimoyama, Supercond. Sci. Technol. **22**, 075008 (2009).

[6] Y. Mizuguchi, F. Tanioka, S. Tsuda, T. Yamaguchi, and T. Takano, Appl. Phys. Lett. **93**, 152505 (2008).

[7] S. Margadonna, Y. Takabayashi, Y. Ohishi, Y. Mizuguchi, Y. Takano, T. Kagayama, T. Nakagawa, M. Takata, and K. Prassides, Phys. Rev. B **80**, 064506 (2009).

[8] S. Medvedev, T. M. McQueen, I. A. Troyan, T. Palasyuk, M. I. Eremets, R. J. Cava, S. Naghavi, F. Casper, V. Ksenofontov, G. Wortmann, and C. Felser: Nat. Mater. **8**, 630 (2009).

[9] E. Bellingeri, I. Pallecchi, R. Buzio, A. Gerbi, D. Marrè, M. R. Cimberle, M. Tropeano, M. Putti, A. Palenzona, and C. Ferdeghini, Appl. Phys. Lett. **96**, 102512 (2012).

[10] P. Mele, K. Matsumoto, Y. Haruyama, M. Mukaida, Y. Yoshida, Y. Ichino, T. Kiss, and A. Ichinose, Supercond. Sci. and Technol. **23**, 052001 (2010).

[11] M.K. Wu, F.C. Hsu, K.W. Yeh, T.W. Huang, J.Y. Luo, M.J. Wang, H.H. Chang, T.K. Chen, S.M. Rao, B.H. Mok, C.L. Chen, Y.L. Huang, C.T. Ke, P.M. Wu, A.M. Chang, C.T. Wu, and T.P. Peng, Physica C **469**, 340 (2009).

[12] Y. Han, W. Y. Li, X. Cao, S. Zhang, B. Xu, and B.R. Zhao, J. Phys, Condens. Matter.

**21**, 235702 (2009).

[13] E. Bellingeri, R. Buzio, A. Gerbi, D. Marrè, S. Congiu, M.R. Cimberle, M. Tropeano, A.S. Siri, A. Palenzona, and C. Ferdeghini, Supercond. Sci. Technol. **22**, 105007 (2009).

[14] W. Si, Z.W. Liy, Q. Jie, W.Q. Yin, J. Zhou, G. Gu, P.D. Jonson, and Q. Li, Appl. Phys. Lett. **95**, 052504 (2009).

[15] Y. Imai, R. Tanaka, T. Akiike, M. Hanawa, I. Tsukada, and A. Maeda, Jpn. J. Appl. Phys. **49**, 023101 (2010).

[16] Y. Imai, T. Akiike, M. Hanawa, I. Tsukada, A. Ichinose, A. Maeda, T. Hikage, T. Kawaguchi, and H. Ikuta, Appl. Phys. Express **3**, 043102 (2010).

[17] M. Hanawa, A. Ichinose, S. Komiya, I. Tsukada, T. Akiike, Y. Imai, T. Hikage, T. Kawaguchi, H. Ikuta, and A. Maeda, Jpn. J. Appl. Phys. **50**, 053101 (2011).

[18] I. Tsukada, M. Hanawa, T. Akiike, F. Nabeshima, Y. Imai, A. Ichinose, S. Komiya, T. Hikage, T. Kawaguchi, H. Ikuta, and A. Maeda, Appl. Phys. Express **4**, 053101 (2011).

[19] B. C. Sales, A. S. Sefat, M. A. McGuire, R. Y. Jin, D. Mandrus and Y. Mozharivskyj, Phys. Rev. B **79**, 094521 (2009).

[20] N. Senguttuvan, M. Aoshima, K. Sumiya, and H. Ishibashi, J. Cryst. Growth **280**, 462 (2005).

[21] M. H. Fang, H. M. Pham, B. Qian, T. J. Liu, E. K. Vehstedt, Y. Liu, L. Spinu, and Z. Q. Mao, Phys. Rev. B **78**, 224503 (2008).

[22] J. D. Jorgensen, B. W. Veal, A. P. Paulikas, L. J. Nowicki, G. W. Crabtree, H. Claus, and W. K. Kwok, Phys. Rev. B **41**, 1863 (1990).

Table I. Sample specifications and growth conditions for films A, B, C and D.

|  | Film A | Film B | Film C | Film D |
|---|---|---|---|---|
| Composition of film | FeTe$_{0.5}$Se$_{0.5}$ | FeTe$_{0.6}$Se$_{0.4}$ | FeTe$_{0.6}$Se$_{0.4}$ | FeTe$_{0.6}$Se$_{0.4}$ |
| Substrate | CaF$_2$ | CaF$_2$ / CaF$_2$ buffer | CaF$_2$ / FeSe buffer | LaAlO$_3$ / CaF$_2$ buffer |
| Deposition Temperature of film / (buffer) [ºC] | 280 | 280 / (340) | 280 / (280) | 280 / (340) |
| Thickness of film / (buffer) [nm] | 115 | 200 / (75) | 130 / (170) | 165 / (60) |
| $T_C^{onset}$ [K] | 16.6 | 13.7 | 14.4 | 13.0 |
| $a$-axis length [nm] | 0.376 | 0.377 | 0.374 | 0.375 |
| $c$-axis length [nm] | 0.594 | 0.605 | 0.604 | 0.605 |
| $c / a$ | 1.579 | 1.604 | 1.615 | 1.611 |

**Figure Captions**

FIG. 1: (color online). Temperature dependence of the resistivities of films A, B, C, and D. Inset: the expanded figure below 20K.

FIG. 2: (color online). Cross-sectional STEM image (a), high-resolution TEM images near the interface region (b) and SAED (c) for film A. The yellow square area shown in Fig. 2 (a) is magnified in Fig. 2 (b). The interface is indicated with a yellow-dashed line in Fig. 2 (b). The marks "F" and "S" in the SAED pattern correspond to the film and substrate, respectively.

FIG. 3: (color online). Cross-sectional STEM image (a), high-resolution TEM images near two interface regions (b) and (c) and SAED (d) for films B. The yellow square areas I and II shown in Fig. 3 (a) are magnified in Fig. 3 (b) and (c), respectively. The (b) and (c) are the interface between the $CaF_2$ substrate and the $CaF_2$ buffer layer and the $CaF_2$ buffer layer and the $FeTe_{0.6}Se_{0.4}$ superconducting film, respectively. The interfaces are indicated as yellow-dashed lines in Fig. 3 (b) and (c). The marks "F" and "B" in the SAED pattern correspond to the film and buffer layer, respectively.

FIG. 4: (color online). Cross-sectional STEM image (a), high-resolution TEM images near two interface regions (b) and (c), and SAED (d) for films C. The yellow square areas I and II shown in Fig. 4 (a) are magnified in Fig. 4 (b) and (c), respectively. (a) and (b) are the interface between the $CaF_2$ substrate and the FeSe buffer layer and between the FeSe buffer layer and the $FeTe_{0.6}Se_{0.4}$ superconducting film, respectively. The interfaces are indicated with yellow-dashed lines in Fig. 4 (a) and (b). The marks "F" and "B" in the SAED pattern correspond to the film and buffer layer, respectively.

FIG. 5: Cross-sectional TEM Images of the FeSe buffer layer (a) and the Fe(Te,Se) superconducting layer (b), which are the magnified areas III and IV shown in Fig. 4

(a), respectively.

FIG. 6: (color online). Cross-sectional STEM image (a), high-resolution TEM images near two interface regions (b) and (c), and SAED patterns (d) and (e) for films D. The yellow square areas I and II shown in Fig. 6 (a) are magnified in Fig. 6 (b) and (c), respectively. (b) and (c) are the interface between the LaAlO$_3$ substrate and the CaF$_2$ buffer layer and the CaF$_2$ buffer layer and the FeTe$_{0.6}$Se$_{0.4}$ superconducting film, respectively. The interfaces are indicated with yellow-dashed lines in Fig. 6 (b) and (c). Fig. 6 (d) is the SAED pattern at the interface between the LaAlO$_3$ substrate and the CaF$_2$ buffer layer. Fig. 6 (e) is the SAED pattern of the superconducting film. The marks "F", "B" and "S" in the SAED patterns correspond to the film, buffer layer and substrate, respectively.

FIG. 7: A nanobeam electron diffraction pattern at the circle III indicated in Fig. 6(b).

FIG. 8: (color online). EDX analyses of each spot (a), (b), (c) and (e) in films A, B, C and D, respectively. EDX analyses of each spot (d) expanded near the interface between the FeSe buffer layer and the substrate of film C.

FIG. 9: (color online). Dependence of the (a) $T_C^{onset}$, (b) $a$- and (c) $c$-axis lengths of grown films on the ratio of $c$-axis length divided by $a$-axis length, $c/a$. Red and black marks indicate oxide substrates and fluoride substrate, respectively. Each mark surrounded by a circle is this experimental data. The marks surrounded by a red oval indicate the composition of FeTe$_{0.6}$Se$_{0.4}$. The other marks indicate the FeTe$_{0.5}$Se$_{0.5}$.

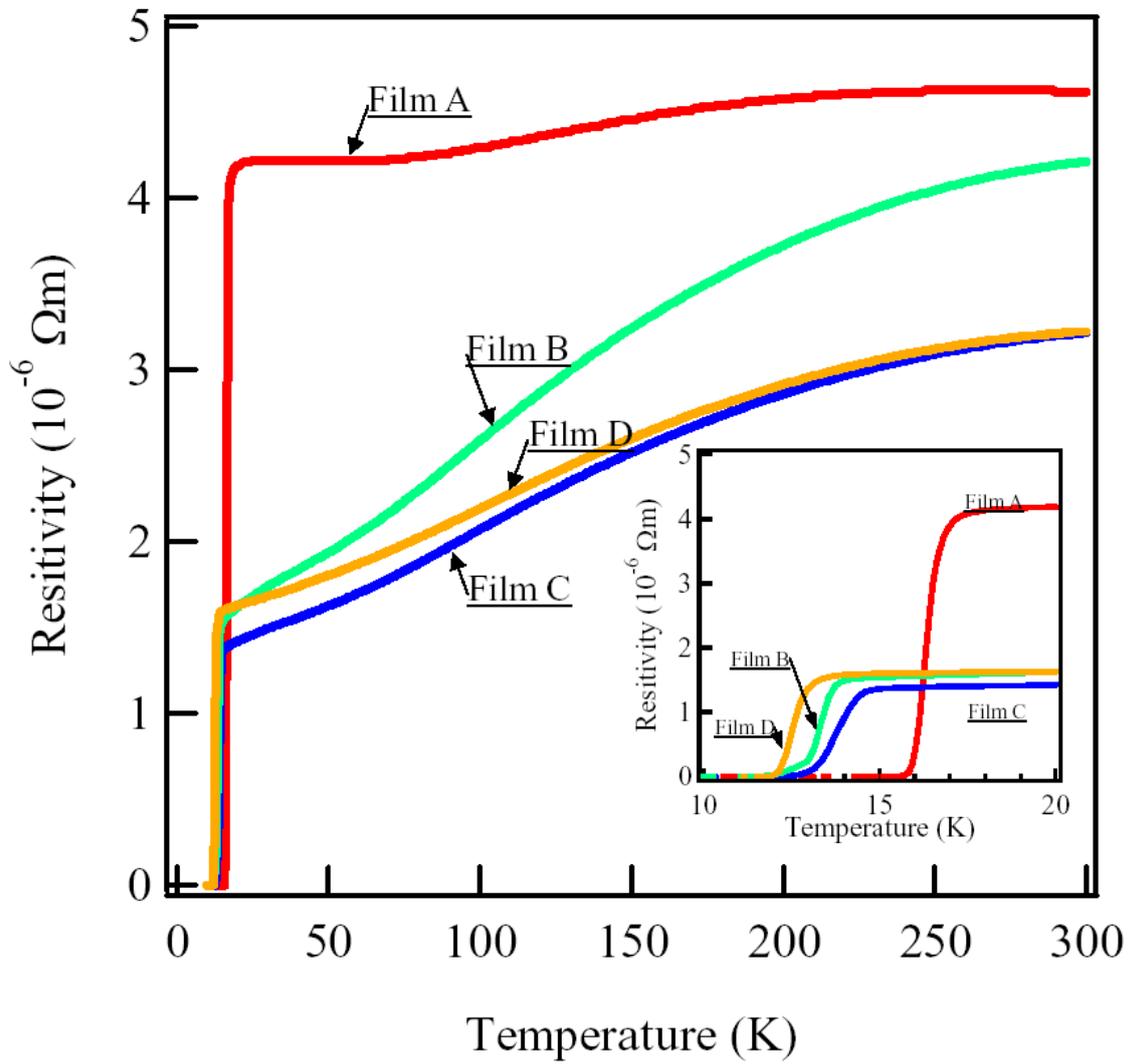

FIG. 1: (color online). Temperature dependence of the resistivities of films A, B, C, and D. Inset: the expanded figure below 20K.

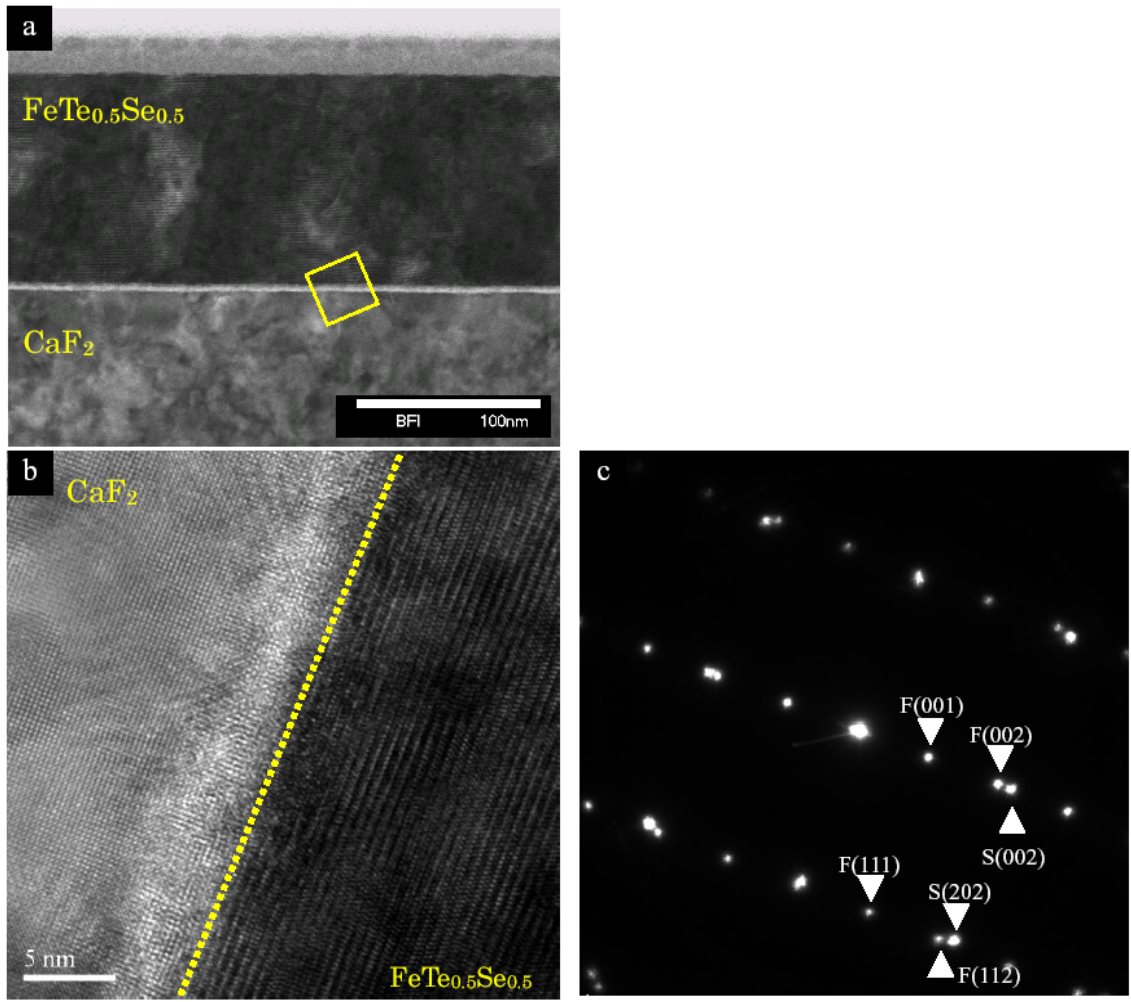

FIG. 2: (color online). Cross-sectional STEM image (a), high-resolution TEM images near the interface region (b) and SAED (c) for film A. The yellow square area shown in Fig. 2 (a) is magnified in Fig. 2 (b). The interface is indicated with a yellow-dashed line in Fig. 2 (b). The marks "F" and "S" in the SAED pattern correspond to the film and substrate, respectively.

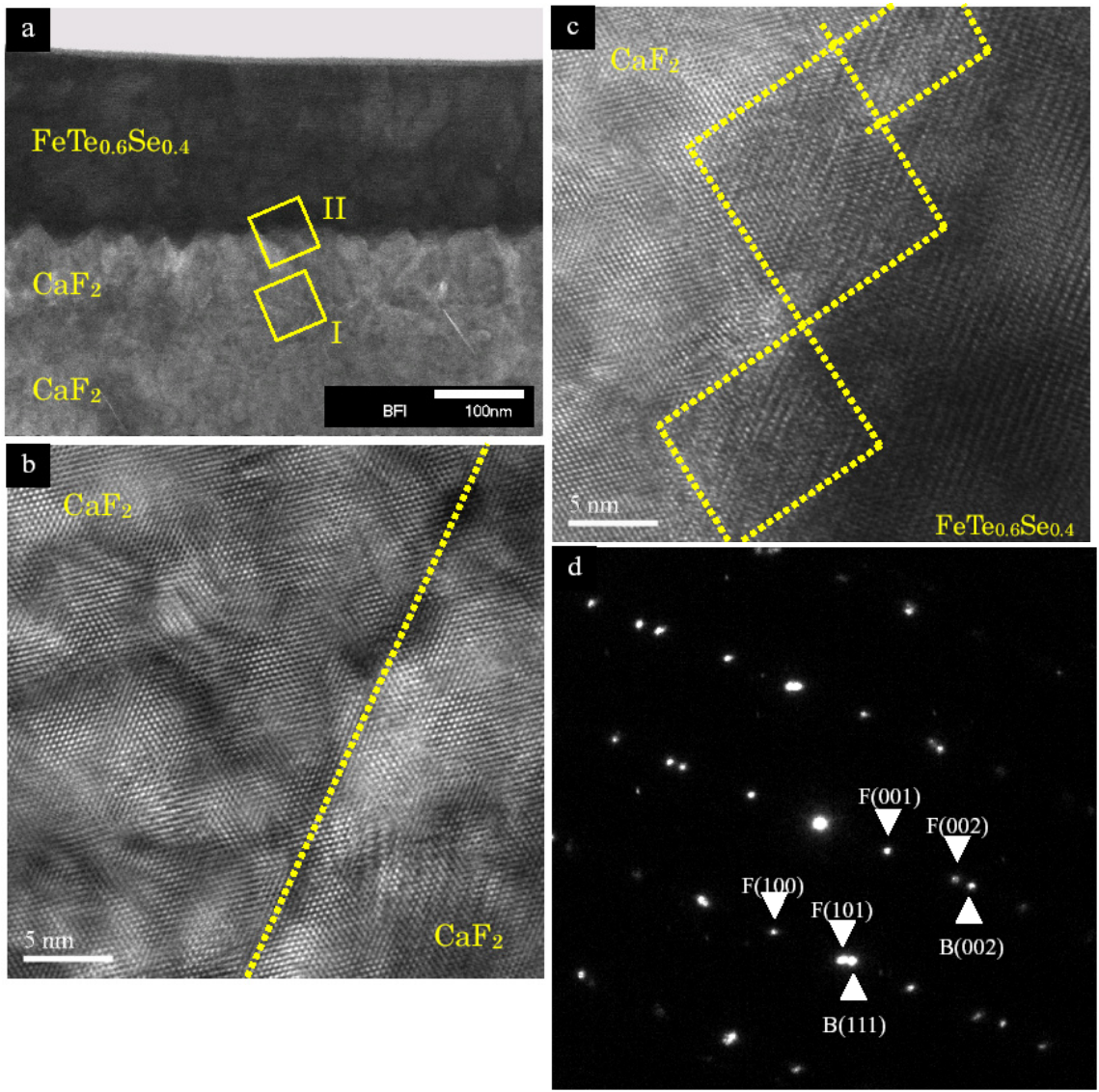

FIG. 3: (color online). Cross-sectional STEM image (a), high-resolution TEM images near two interface regions (b) and (c) and SAED (d) for films B. The yellow square areas I and II shown in Fig. 3 (a) are magnified in Fig. 3 (b) and (c), respectively. The (b) and (c) are the interface between the $CaF_2$ substrate and the $CaF_2$ buffer layer and the $CaF_2$ buffer layer and the $FeTe_{0.6}Se_{0.4}$ superconducting film, respectively. The interfaces are indicated as yellow-dashed lines in Fig. 3 (b) and (c). The marks "F" and "B" in the SAED pattern correspond to the film and buffer layer, respectively.

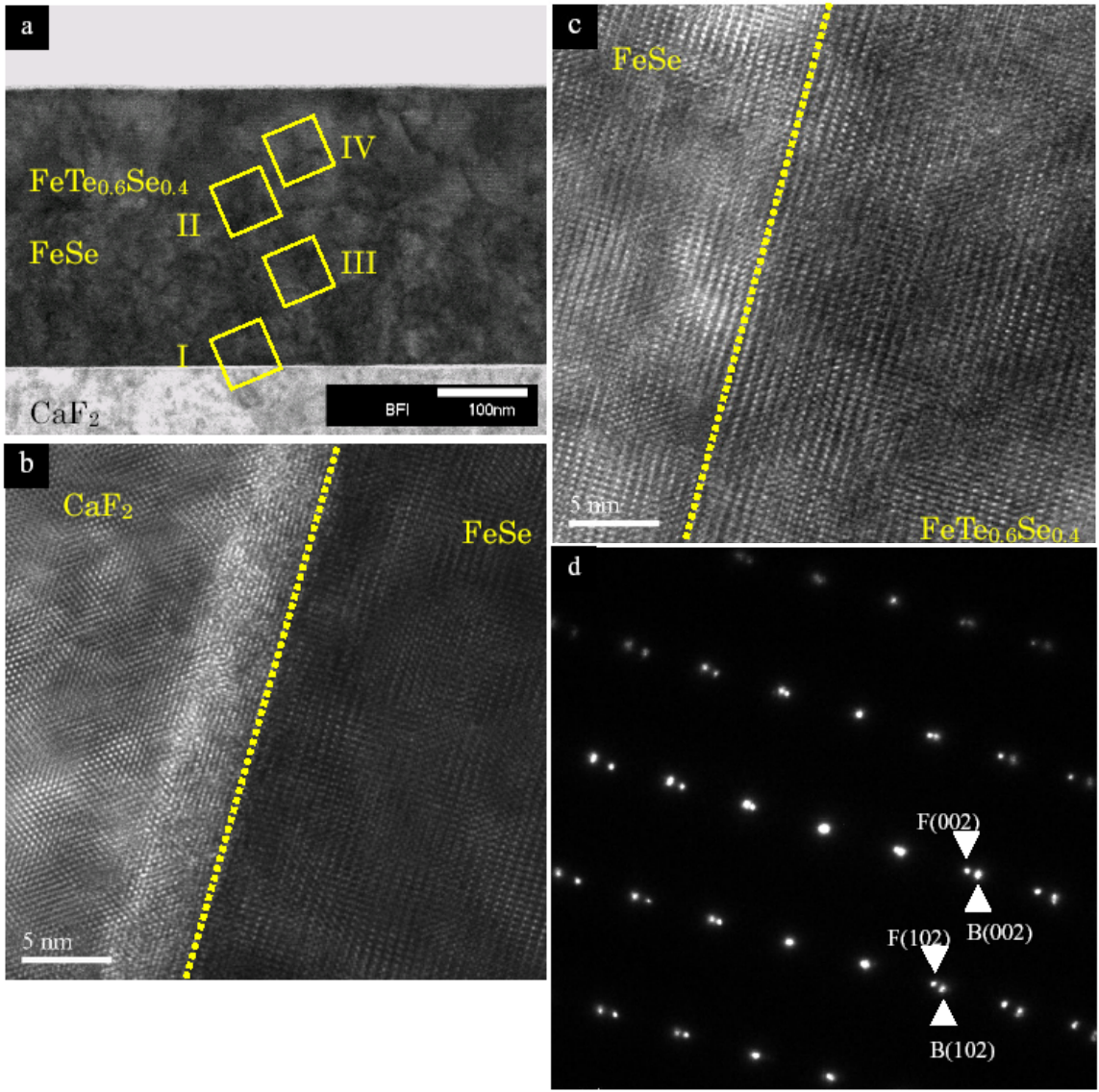

FIG. 4: (color online). Cross-sectional STEM image (a), high-resolution TEM images near two interface regions (b) and (c), and SAED (d) for films C. The yellow square areas I and II shown in Fig. 4 (a) are magnified in Fig. 4 (b) and (c), respectively. (a) and (b) are the interface between the $CaF_2$ substrate and the FeSe buffer layer and between the FeSe buffer layer and the $FeTe_{0.6}Se_{0.4}$ superconducting film, respectively. The interfaces are indicated with yellow-dashed lines in Fig. 4 (a) and (b). The marks "F" and "B" in the SAED pattern correspond to the film and buffer layer, respectively.

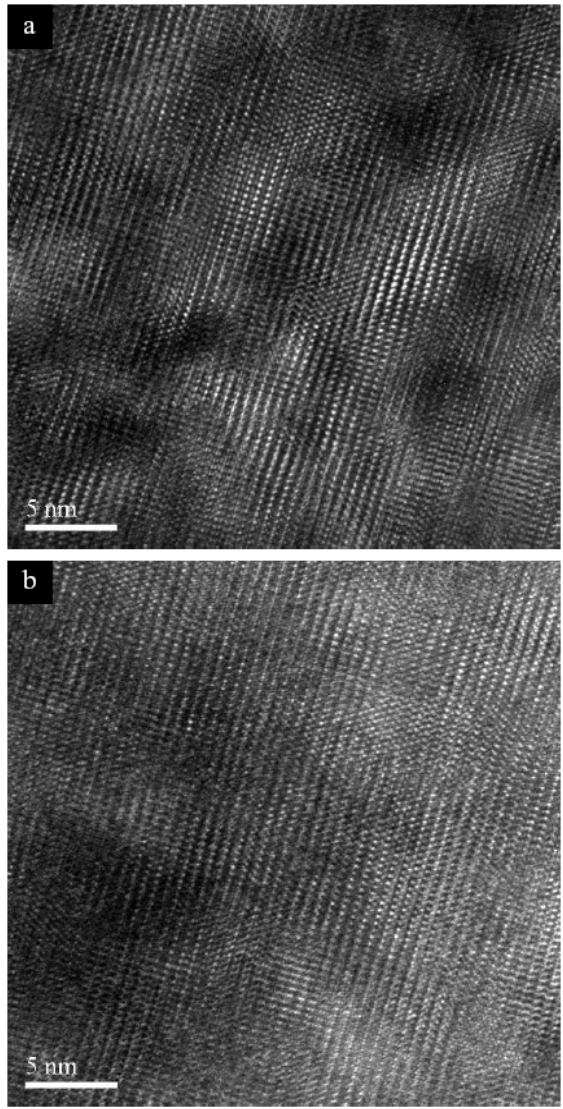

FIG. 5: Cross-sectional TEM Images of the FeSe buffer layer (a) and the Fe(Te,Se) superconducting layer (b), which are the magnified areas III and IV shown in Fig. 4 (a), respectively.

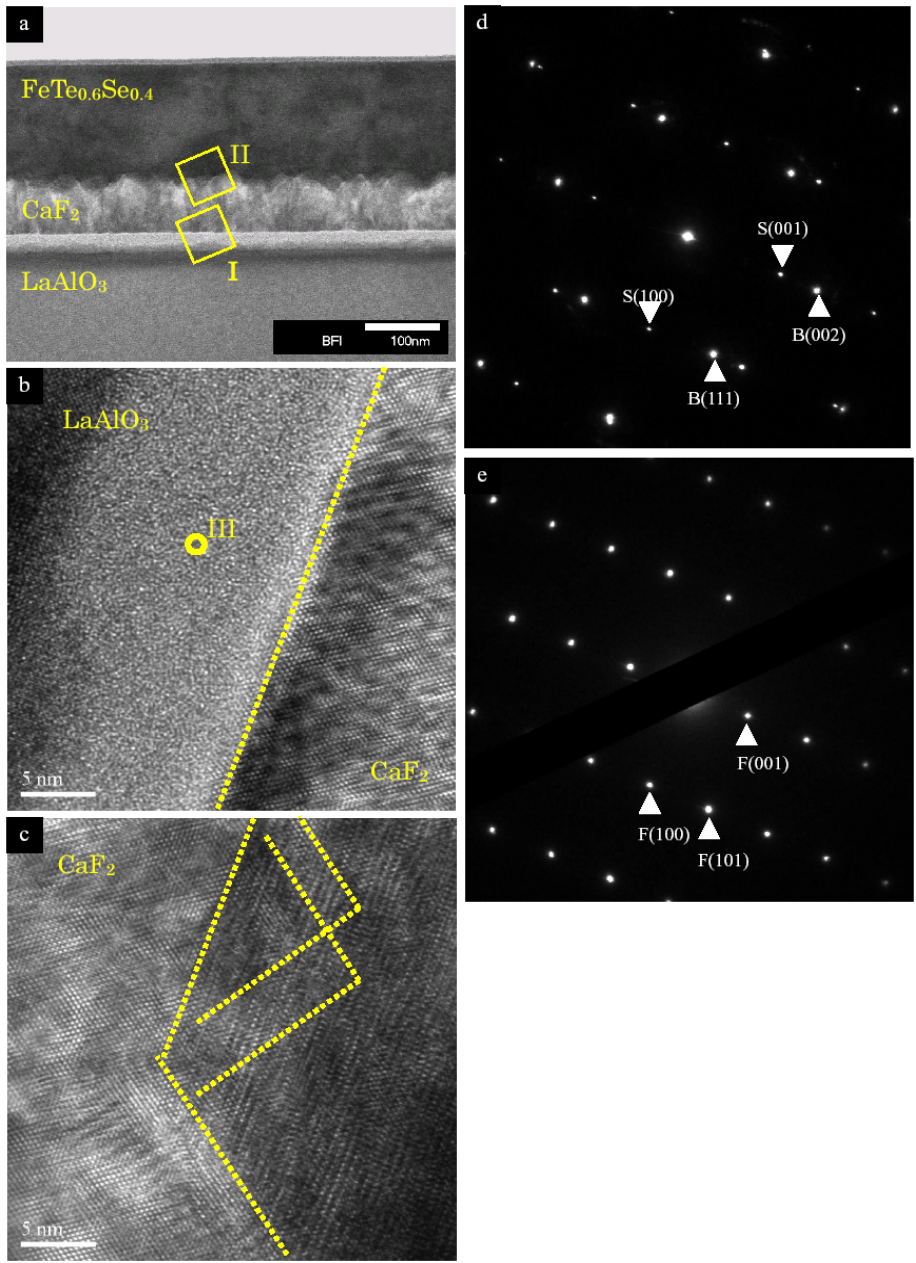

FIG. 6: (color online). Cross-sectional STEM image (a), high-resolution TEM images near two interface regions (b) and (c), and SAED patterns (d) and (e) for films D. The yellow square areas I and II shown in Fig. 6 (a) are magnified in Fig. 6 (b) and (c), respectively. (b) and (c) are the interface between the LaAlO$_3$ substrate and the CaF$_2$ buffer layer and the CaF$_2$ buffer layer and the FeTe$_{0.6}$Se$_{0.4}$ superconducting film, respectively. The interfaces are indicated with yellow-dashed lines in Fig. 6 (b) and (c). Fig. 6 (d) is the SAED pattern at the interface between the LaAlO$_3$ substrate and the CaF$_2$ buffer layer. Fig. 6 (e) is the SAED pattern of the superconducting film. The marks "F", "B" and "S" in the SAED patterns correspond to the film, buffer layer and substrate, respectively.

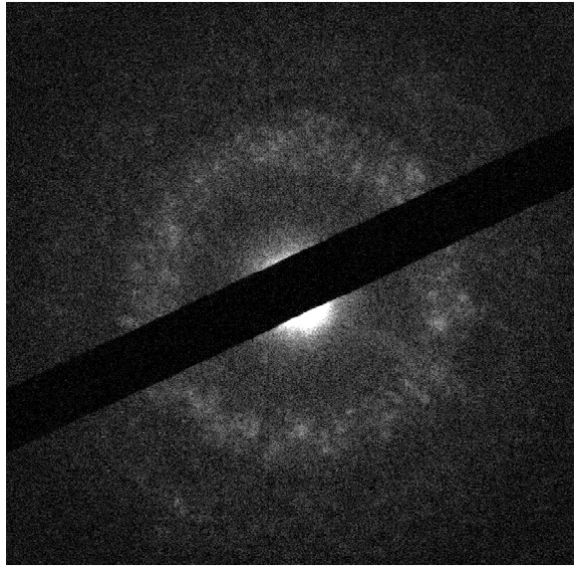

FIG. 7: A nanobeam electron diffraction pattern at the circle III indicated in Fig. 6(b).

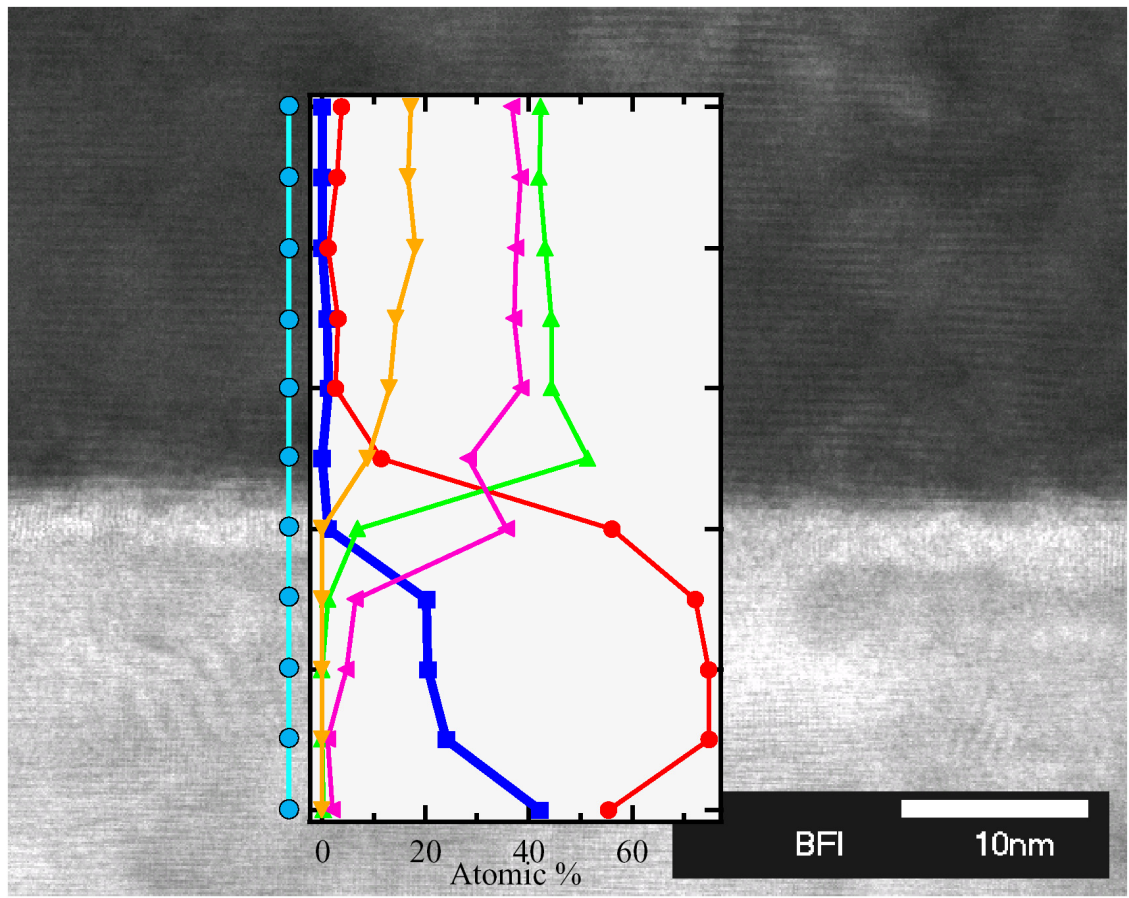

FIG. 8 (a): (color online). EDX analyses of each spot (a) ,(b), (c) and (e) in films A, B, C and D, respectively. EDX analyses of each spot (d) expanded near the interface between the FeSe buffer layer and the substrate of film C.

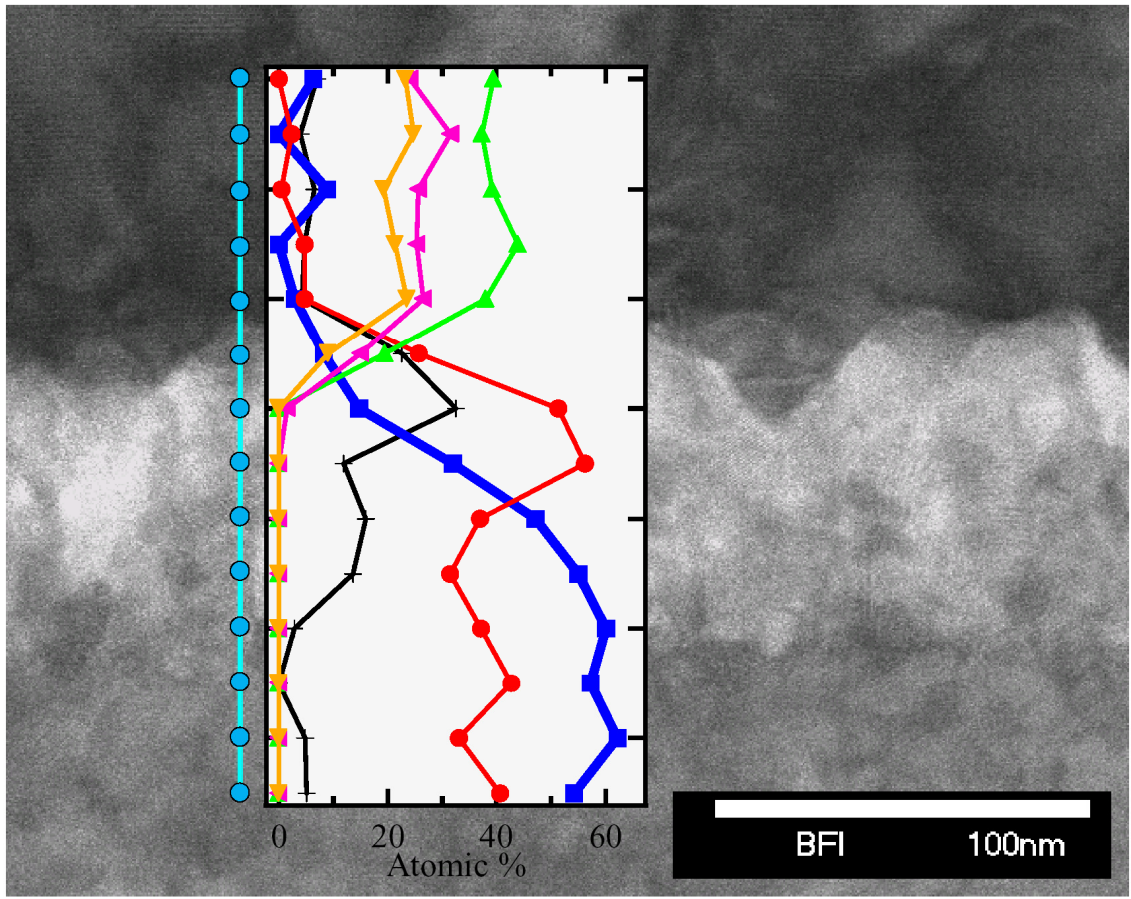

FIG. 8 (b): (color online). EDX analyses of each spot (a) ,(b), (c) and (e) in films A, B, C and D, respectively. EDX analyses of each spot (d) expanded near the interface between the FeSe buffer layer and the substrate of film C.

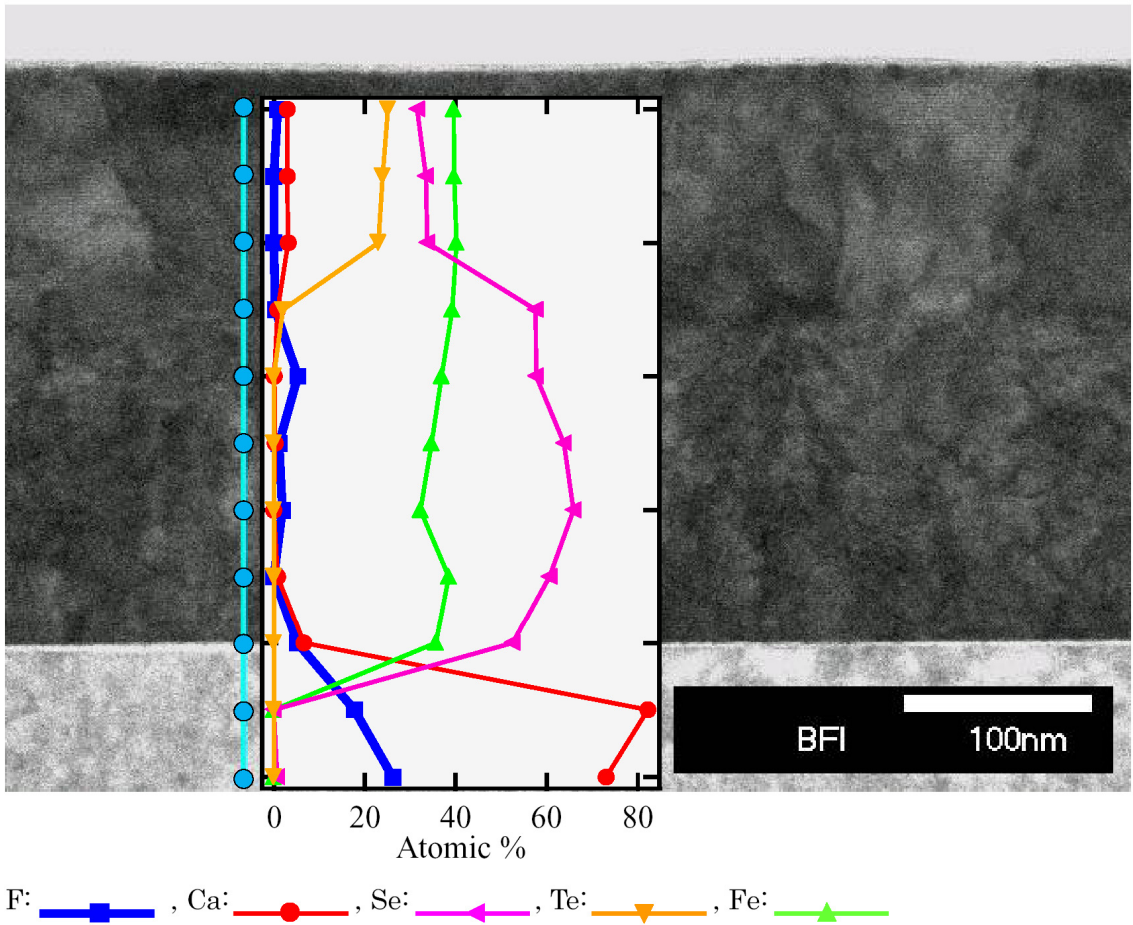

FIG. 8 (c): (color online). EDX analyses of each spot (a) ,(b), (c) and (e) in films A, B, C and D, respectively. EDX analyses of each spot (d) expanded near the interface between the FeSe buffer layer and the substrate of film C.

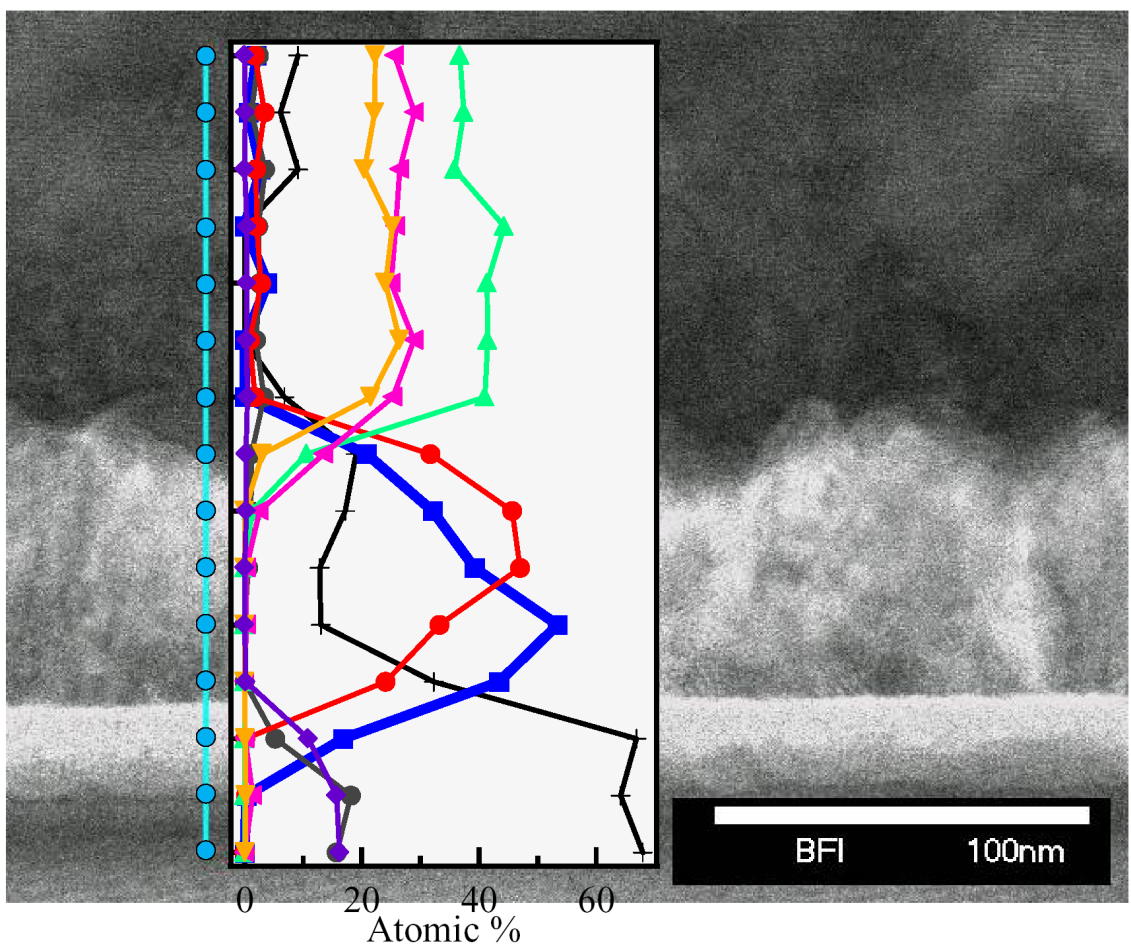

FIG. 8 (d): (color online). EDX analyses of each spot (a) ,(b), (c) and (e) in films A, B, C and D, respectively. EDX analyses of each spot (d) expanded near the interface between the FeSe buffer layer and the substrate of film C.

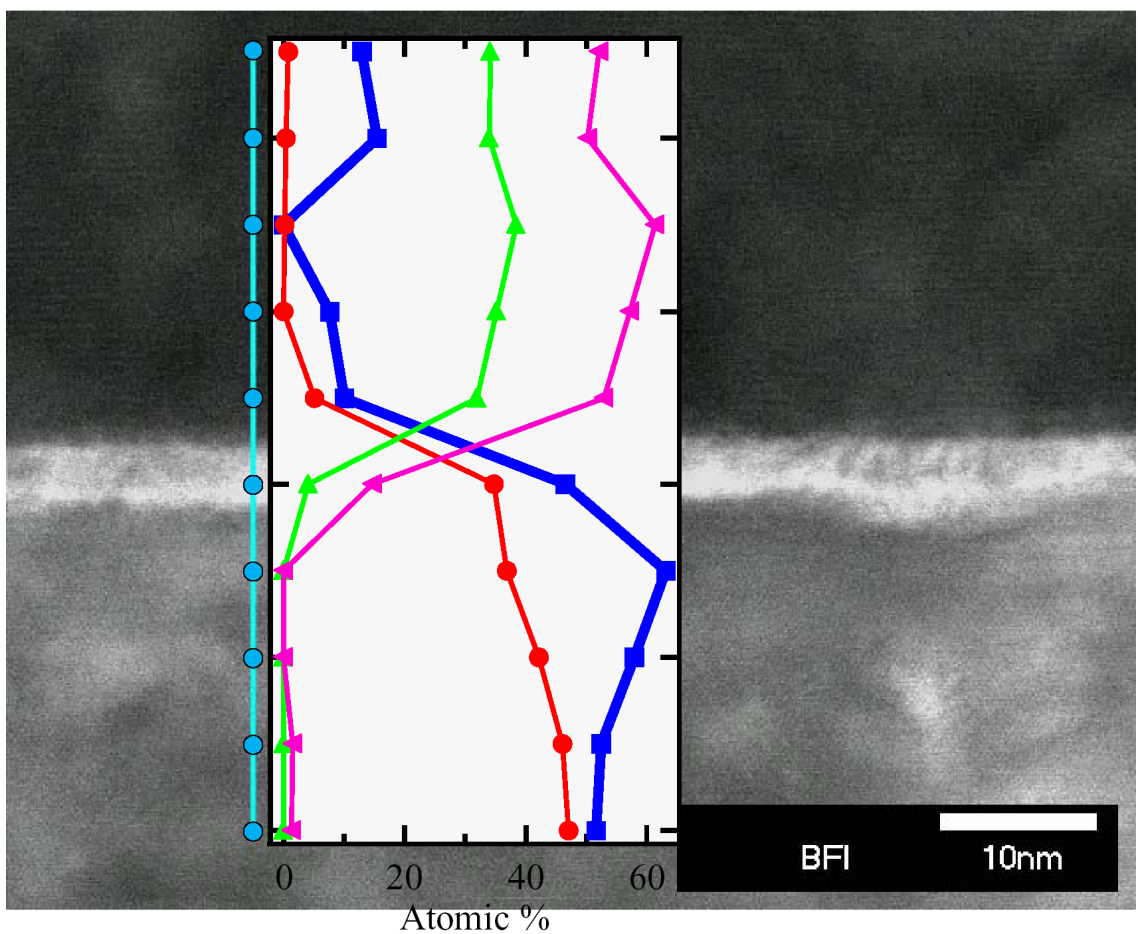

FIG. 8 (e): (color online). EDX analyses of each spot (a) ,(b), (c) and (e) in films A, B, C and D, respectively. EDX analyses of each spot (d) expanded near the interface between the FeSe buffer layer and the substrate of film C.

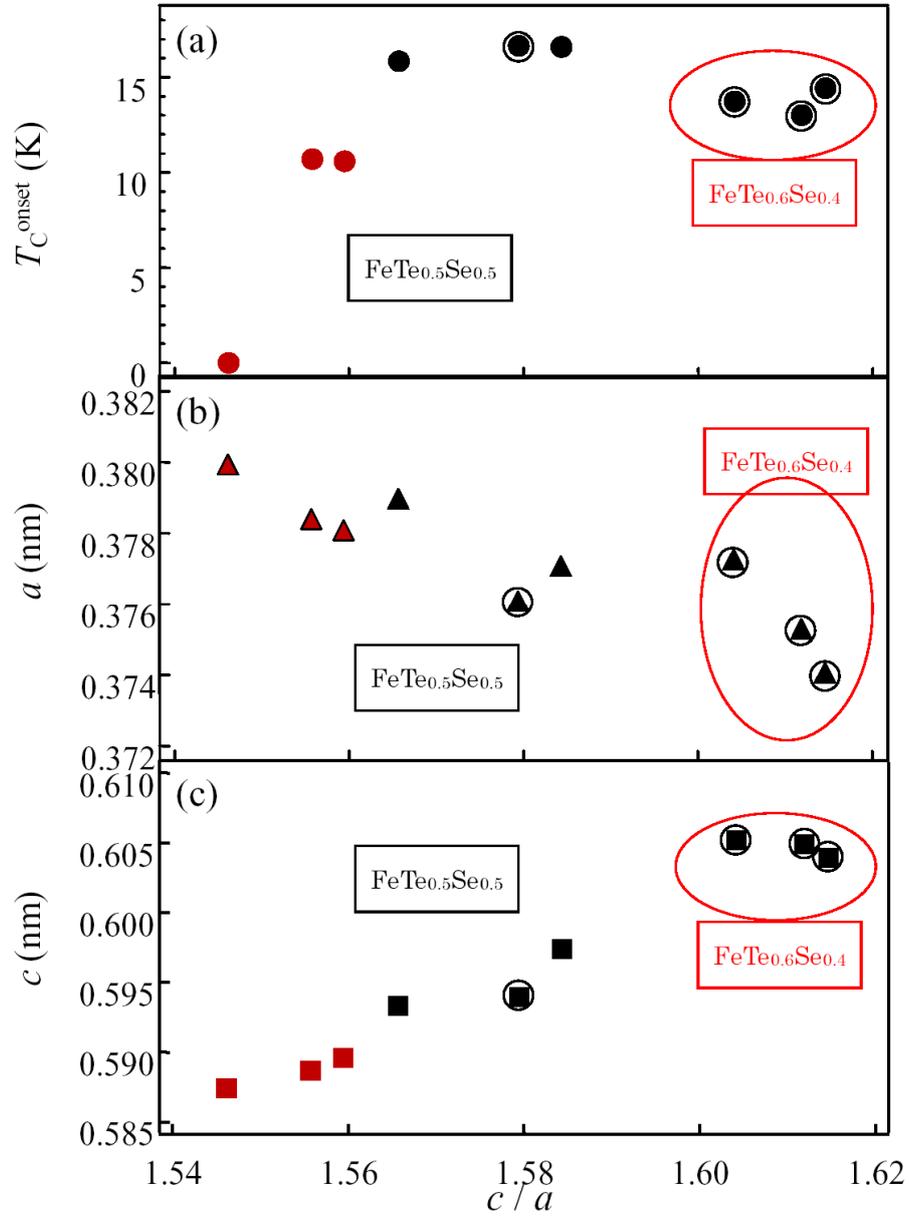

FIG. 9: (color online). Dependence of the (a) $T_C^{onset}$, (b) $a$- and (c) $c$-axis lengths of grown films on the ratio of $c$-axis length divided by $a$-axis length, $c/a$. Red and black marks indicate oxide substrates and fluoride substrate, respectively. Each mark surrounded by a circle is this experimental data. The marks surrounded by a red oval indicate the composition of FeTe$_{0.6}$Se$_{0.4}$. The other marks indicate the FeTe$_{0.5}$Se$_{0.5}$.